\documentclass[a4paper,11pt]{article}
\usepackage{pos}   
\usepackage{svg}

\usepackage{tikz}

\newcommand{\fatg}{{\rm{I}}\!\Gamma}
\begin{document}

\title{General kinematics of the three-gluon vertex from quenched lattice QCD}
%
%

\author*[a]{(HSV\footnote{Huelva-Sevilla-Valencia} Collaboration) F.~Pinto-G\'omez}
\author[a]{F.~De Soto}
\author[b]{M. N. Ferreira}
\author[b]{J. Papavassiliou}
\author[c]{J. Rodríguez-Quintero}

\affiliation[a]{Dpto. Sistemas F\'isicos, Qu\'imicos y Naturales, Univ. Pablo de Olavide, 41013 Sevilla, Spain
}
\affiliation[b]{Department of Theoretical Physics and IFIC, University of Valencia and CSIC, E-46100, Valencia, Spain}
\affiliation[c]{Dpto. Ciencias Integradas, Centro de Estudios Avanzados en Fis., Mat. y Comp., Fac. Ciencias Experimentales, Universidad de Huelva, Huelva 21071, Spain}

\emailAdd{fernandomalaga97@gmail.com}
\emailAdd{fcsotbor@upo.es}
\emailAdd{mnarciso@ifi.unicamp.br}
\emailAdd{Joannis.Papavassiliou@uv.es}
\emailAdd{jose.rodriguez@dfaie.uhu.es}
\abstract{%

We present new results for the transversely projected three-gluon vertex from quenched lattice QCD simulations using standard Wilson action.
While previous works focused in some particular kinematics such as the  symmetric $(q^2 = r^2 = p^2)$ and soft-gluon $(p = 0)$ cases, here we 
will present a detailed analysis of the bisectoral case $(r^2 = q^2 \ne p^2)$ where the transversely projected vertex can be cast in terms of three independent tensors.
The lattice data show a clear dominance of the form-factor corresponding to the tree-level tensor, whose dependence on the momenta can be almost entirely expressed in terms of the symmetric combination of the momenta $s^2=(q^2+r^2+p^2)/2$.

}
\FullConference{%
  The 39th International Symposium on Lattice Field Theory\\
  8-13 August 2022\\
  Bonn, Germany
}

\maketitle
\section{Introduction}
\label{sec:intro}

Despite the apparent simplicity of the QCD Lagrangian, there is an enormous wealth of emergent phenomena that arise dynamically~\cite{Roberts:2020hiw}. Already at the level of gluon dynamics, Yang-Mills theory is massless at the level of the Lagrangian, and remain so at any order in perturbation theory, but a gluon mass is dynamically generated, a phenomenon that has achieved a noticeable consensus over the last decades from both lattice and Schwinger-Dyson Equations (DSE) ~\cite{Cucchieri:2006tf,Cucchieri:2008qm,Huber:2016tvc,Huber:2012zj,Pelaez:2013cpa,Aguilar:2013vaa,Blum:2014gna,Mitter:2014wpa,Williams:2015cvx,Blum:2015lsa,Cyrol:2016tym,Duarte:2016ieu,Sternbeck:2017ntv,Corell:2018yil,Vujinovic:2018nqc,Barrios:2022hzr}. 

The intricate non-perturbative dynamics of Yang-Mills theory has attracted the attention of lattice and Dyson-Schwinger equations (DSE) practitioners for decades, and among the gluon Green functions, the three gluon vertex constitutes a fundamental part of Yang-Mills dynamics. It is a key ingredient in many phenomenological continuum analysis and has been related to the mechanism of gluon mass generation~\cite{Aguilar:2019kxz,Aguilar:2021uwa,Pinto-Gomez:2022brg}. The lattice non-perturbative evaluation of the three-gluon vertex in soft-gluon and symmetric kinematics has found an infrared suppression \cite{Athenodorou:2016oyh,Boucaud:2017obn} which has been related to the fact that the ghost remains massless while the gluon dynamically acquires a mass \cite{Aguilar:2021lke,Aguilar:2021okw}.

\section{Kinematics of the three-gluon vertex}
\begin{figure}[t]
	\begin{center}
		\begin{tabular}{ccc}
				\includegraphics[scale=0.35]{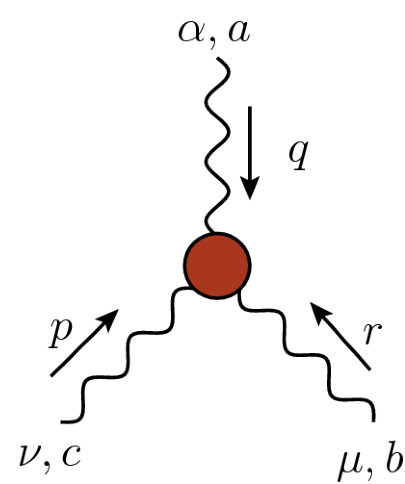} 
			&		
			    \begin{tikzpicture}[scale=0.4]
					\fill[fill=blue!30] (0,3) -- (3,0) -- (-1.25,-1.25) -- (0,3);
					\draw [very thick, -latex] (0,0) -- (0,5) node [right] {$p^2$};
					\draw [very thick, -latex] (0,0) -- (-2.5,-2.5) node [right] {$q^2$};
					\draw [very thick, -latex] (0,0) -- (5,0) node [right] {$r^2$};
					\draw [thick, blue] (0,3) -- (3,0) -- (-1.25,-1.25) -- (0,3);
					\draw [thick, blue, -latex] (0,0) -- (3.5,3.5);
					\node at (0.75,0.75) [circle,fill,inner sep=1.2pt,green]{};
				\end{tikzpicture}
			&
			\begin{tikzpicture}[scale=0.5]
					\node (r) at ( 3.0,  0.0) {}; %
					\node (q) at (-3.0,  0.0) {}; %
					\node (p) at ( 0.0, 5.36) {}; %
					\node (p2) at ( 0.0, 3.55) {}; 
					\node (S) at (0,1.75) {};
					\node (Op) at (0,2.625) {};
					\node (Oq) at (-0.76,1.31) {};
					\node (Or) at (0.76,1.31) {};
					\node (SGp) at (0,0) {};
					\node (SGq) at (1.52,2.63) {};
					\node (SGr) at (-1.52,2.63) {};
					\node (apA) at ( 3.0,  1.75) {}; %
					\node (amA) at ( -3.0,  1.75) {}; %
					\node (apB) at ( 0,  6.25) {};
					\fill[fill=blue!20] (p.center) -- (r.center) -- (q.center) -- (p.center);	-
					\draw [thick, blue] (r) -- (q) -- (p) -- (r);
					\fill[fill=white] (S) circle (1.72cm);
					\draw [very thick,gray] (p2) -- (SGp);
					\node at (-0.4,2.05) {$S$};
					\node at (S)[circle,fill,inner sep=1.5pt,green]{};
					\node at (0,-0.6) {${p^2=0}$};
					\node at (SGp) [circle,fill,inner sep=1.5pt,orange]{};
					\node at (2.8,2.8) {${q^2=0}$};
					\node at (SGq) [circle,fill,inner sep=1.5pt,black]{};
					\node at (-2.8,2.8) {${r^2=0}$};
					\node at (SGr) [circle,fill,inner sep=1.5pt,black]{};
				\end{tikzpicture} \\
				(a) & (b) & (c) 
			\end{tabular}
	\end{center}
	\caption{The kinematic configuration of the three-gluon vertex (left diagram) represented by the Cartesian coordinates $(q^2,r^2,p^2)$ (right picture). All kinematics allowed are contained in a circle around the symmetric case ($q^2=r^2=p^2$, green dot). The bisectoral line (thick gray), and the particular soft-gluon (orange solid circle), and symmetric (green) cases appear depicted. The other two soft-gluon limits (black) are also illustrated.  
	}
	\label{fig:triangle}
\end{figure}
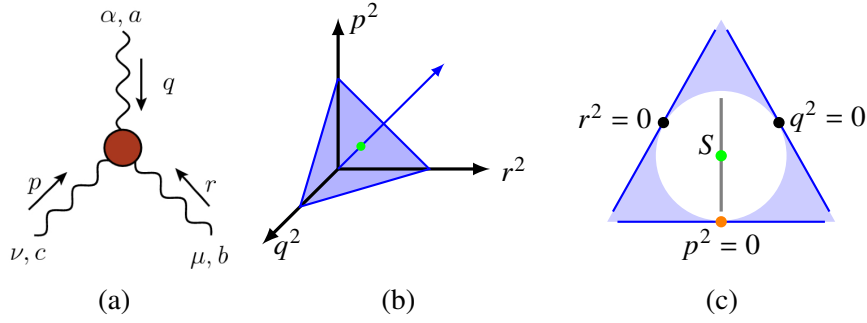

Denoting the  momenta of the three incoming gluons as $q$, $r$ and $p$ (see Fig.\ref{fig:triangle} (a)) any scalar form factor of the three-gluon vertex can only depend on the squared momenta $q^2$, $r^2$ and $p^2$. While previous studies focused on the more restricted particular cases $q^2=r^2=p^2$ (dubbed symmetric case) and $p=0$ (and thus $q^2=r^2$, called soft-gluon case), we will extend this analysis to the case $q^2=r^2$, and thus the form factors depend only on two momenta $q^2$ and $p^2$ or, alternatively, one momenta and one angle, as for example the angle between $q$ and $r$, $\theta_{qr}$, related to the squared momenta through:
\begin{equation}
    \cos \theta_{qr}  = \frac{p^2-q^2-r^2}{2 \sqrt{q^2 r^2}}\ .
\end{equation}
This angle takes values between $0$ and $\pi$ for this bisectoral case ($q^2=r^2$), with the special kinematics mentioned above, corresponding to  $\theta_{qr}=2\pi/3$ for the symmetric case and $\theta_{qr}=\pi$ for the $p=0$ soft-gluon case, being both particular kinematical cases of the bisectoral one.

Any kinematic configuration defined by the values of $q^2$, $r^2$ and $p^2$ can be represented in the trihedron of Fig.~\ref{fig:triangle}(b), with the possible values of these variables subject to the condition $q+r+p=0$ restricted to a cone whose axis is the $q^2=r^2=p^2$ blue arrow represented in Fig.~\ref{fig:triangle}(b). The possible kinematics are best illustrated by restricting to a constant $q^2+r^2+p^2$ plane (blue triangle), where the allowed kinematics are contained in its incircle (white circle in Fig.\ref{fig:triangle}(c). Here we will focus in the dubbed bisectoral kinematic configurations, $q^2=r^2$, represented by the grey line in Fig~\ref{fig:triangle}(c), where also the symmetric and soft-gluon cases have been depicted.

\section{Transversely projected vertex}
\label{sec:vertex}

Being $\widetilde{A}_\mu^a (q) $ the gluon field in momentum-space in Landau gauge as obtained from lattice-QCD, we compute the three-point correlation function $\langle\widetilde{A}_\alpha^a (q)\widetilde{A}_\mu^b (r)\widetilde{A}_\nu^c (p)\rangle$. Focusing in the color anti-symmetric contribution, we defined the quantity
\begin{equation}
   \mathcal{G}_{\alpha\mu\nu} (q,r,p)= \frac{1}{24} f_{a b c} \langle\widetilde{A}_\alpha^a (q)\widetilde{A}_\mu^b (r)\widetilde{A}_\nu^c (p)\rangle
\end{equation}
which contains the information of the transversely projected three-gluon vertex, $\overline{\Gamma}_{\alpha\mu\nu} (q,r,p)$. A couple of comments are in order here: first, the lattice only has  access to the transverse part of the three-gluon vertex, and not the full one-particle irreducible (1PI) vertex $\fatg_{\alpha\mu\nu}$. Second, and as a consequence of that, any term comprising massless longitudinal poles $V_{\alpha\mu\nu}$ (such as those entering the Schwinger mechanism for gluon mass generation~\cite{Ibanez:2012zk,Binosi:2017rwj,Aguilar:2021uwa} cannot be directly accessed. In other words, if the 1PI vertex can be decomposed as $\fatg_{\alpha\mu\nu}=\Gamma_{\alpha\mu\nu} + V_{\alpha\mu\nu}$, with $\Gamma_{\alpha\mu\nu}$ the pole-free component,
only the pole-free transversely projected vertex $\overline\Gamma_{\alpha\mu\nu}(q,r,p)=\Gamma_{\alpha'\mu'\nu'}(q,r,p) P_{\alpha}^{\alpha'}(q) P_{\mu}^{\mu'}(r) P_{\nu}^{\nu'}(p)$ can be directly computed from lattice-QCD.

From $\mathcal{G}_{\alpha\mu\nu} (q,r,p)$, one  can extract the transversely projected vertex as:
\begin{equation}
    g \overline{\Gamma}_{\alpha\mu\nu} (q,r,p) = \frac{\mathcal{G}_{\alpha\mu\nu} (q,r,p)}{\Delta(q^2)\Delta(r^2)\Delta(p^2)}
\end{equation}
As it corresponds to the transverse projection of the pole-free vertex, it can be cast in terms of the transverse tensors of the full Ball-Chiu decomposition \cite{Ball:1980ax,Ball:1980ay} of the three-gluon vertex. Instead, we will construct a basis of transverse tensors for the general three-gluon vertex which are anti-symmetric under Bose transformations $\tilde\lambda_i \to - \tilde{\lambda}_i$ according to 
\begin{subequations}
\label{eq:tensors}
\begin{align}
\tilde\lambda_1^{\alpha \mu \nu} \!=& \ P_{\alpha'}^\alpha(q)  P_{\mu'}^{\mu}(r)  P_{\nu'}^\nu(p) 
\left[\ell_1^{\alpha'\mu'\nu'} + \ell_4^{\alpha'\mu'\nu'} + \ell_7^{\alpha'\mu'\nu'}\right] \,,
\label{eq:tl1}
\\ 
\tilde\lambda_2^{\alpha \mu \nu} \!=&  \frac{3}{2 s^2} \,(q-r)^{\nu'} (r-p)^{\alpha'} (p-q)^{\mu'} 
P_{\alpha'}^\alpha(q)  P_{\mu'}^{\mu}(r)  P_{\nu'}^\nu(p)\,,
\label{eq:tl2}
\\
\tilde\lambda^{\alpha \mu \nu} \!=& \frac{3}{2 s^2}  P_{\alpha'}^\alpha(q)  P_{\mu'}^{\mu}(r)  P_{\nu'}^\nu(p) 
\left[\ell_3^{\alpha'\mu'\nu'} + \ell_6^{\alpha'\mu'\nu'} + \ell_9^{\alpha'\mu'\nu'}\right]\,,
\label{eq:tl3}
\\
\tilde\lambda_4^{\alpha \mu \nu} \!=& \left( \frac{3}{2 s^2}\right)^2
\left[t_1^{\alpha\mu\nu} + t_2^{\alpha\mu\nu} + t_3^{\alpha\mu\nu}\right]\,,
\label{eq:tl4}
\end{align}
\end{subequations}
where $l_i$ and $t_i$ are the tensors from Ball-Chiu decomposition \cite{Ball:1980ax,Ball:1980ay} and the first tensor corresponds to the tree-level tensor, and both the first and second tensors coincide in the symmetric limit with the tensors used in previous works \cite{Boucaud:2018xup,Aguilar:2021okw}. For the bisectoral case  ($q^2=r^2$) we will keep $\lambda_i = \lim_{r^2\to q^2} \tilde\lambda_i$ for $i=1,\cdots,3$, $\lambda_4$ being a linear combination of the rest. Using this basis we will write:
\begin{equation}
    \overline{\Gamma}_{\alpha\mu\nu} (q,r,p) = \sum_i \overline{\Gamma}_i(q^2,r^2,p^2) \lambda^i_{\alpha\mu\nu} (q,r,p)\,.
\end{equation}
where $\overline{\Gamma}_i(q^2,r^2,p^2)$ are the scalar form factors. Note that with the use of this basis, the form factors do not change when any pair of momenta are interchanged, i.e., $\overline{\Gamma}_i(q^2,r^2,p^2)=\overline{\Gamma}_i(q^2,p^2,r^2)=\cdots$. Consequently, they can only depend on Bose-symmetric combinations of momenta.

The scalar form factors $\overline{\Gamma}_i(q^2,r^2,p^2)$ can be extracted from the lattice data for $\overline{\Gamma}_{\alpha\mu\nu} (q,r,p)$ by solving the linear system:
\begin{equation}\label{eq:linearsystem}
    \underbrace{\overline{\Gamma}_{\alpha\mu\nu} (q,r,p) \cdot \lambda_j^{\alpha\mu\nu}(q,r,p)}_{b_j} = \sum_i \overline{\Gamma}_i(q^2,r^2,p^2) \underbrace{\lambda^i_{\alpha\mu\nu} (q,r,p) \lambda_j^{\alpha\mu\nu}(q,r,p)}_{M_{ij}}
\end{equation}
or, calling $\mathcal{P}_i^{\alpha\mu\nu}=M^{-1}_{ij} \lambda_j^{\alpha\mu\nu}(q,r,p)$ the projector over each element of the basis:
\begin{equation}
\label{eq:projectors}
    \overline{\Gamma}_i(q^2,r^2,p^2) = \mathcal{P}_i^{\alpha\mu\nu} \overline{\Gamma}_{\alpha\mu\nu} (q,r,p)\,.
\end{equation}

When working in the bisectoral case, some kinematics are close to the symmetric or soft-gluon cases, where the three tensors in Eq.(\ref{eq:tensors}) are not linearly independent, and hence, the determinant of the matrix $M$ in Eq.(\ref{eq:linearsystem}) will be close to  zero. These kinematical points present a rather pronounced increase of the noise, and are discarded in the following.

\section{Results}
\label{sec:results}

\begin{figure}[t]
     \includegraphics[width=0.98\columnwidth]{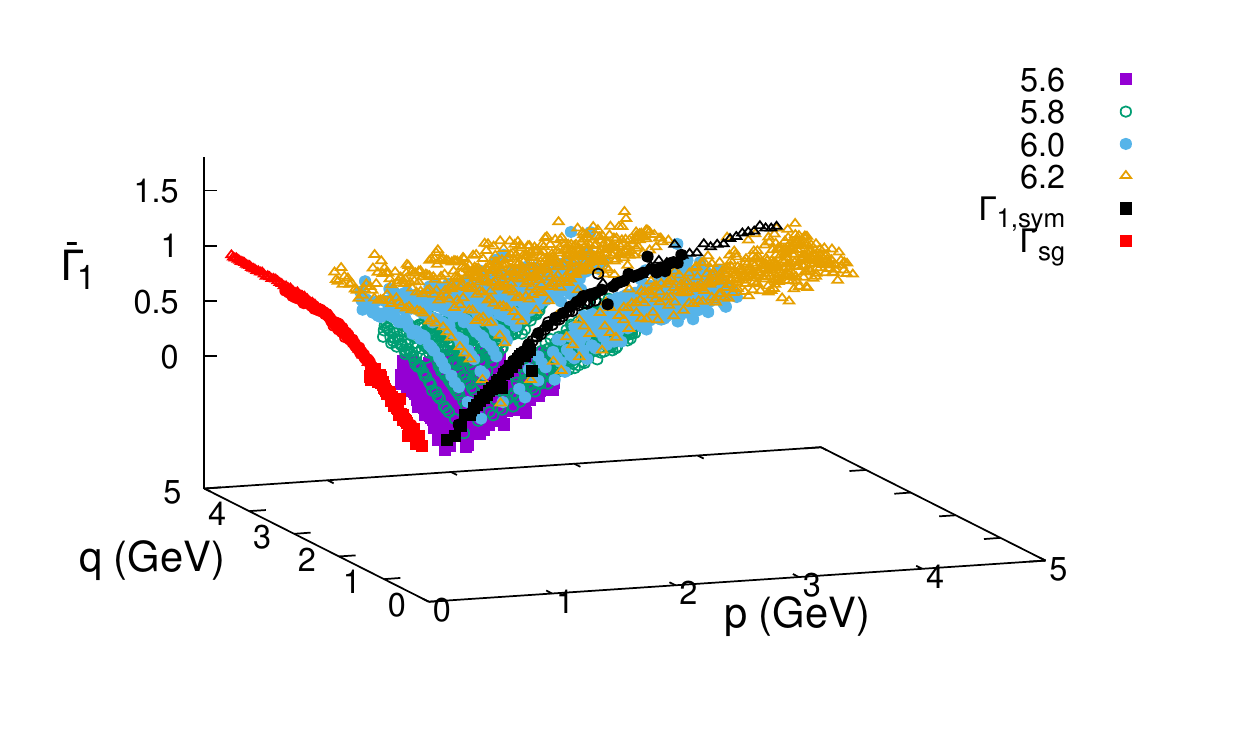} 
     \vspace{-1.6cm}
\caption{Three-gluon form factor for the tree-level tensor $\lambda_1(q,r,p)$ for bisectoral kinematic configurations $q^2=r^2$ as a function of the momenta $q=\sqrt{q^2}=\sqrt{r^2}$ and $p=\sqrt{p^2}$ for four different values of $\beta$ (in the figure caption). Soft-gluon case (corresponding to $p=0$, red symbols) and symmetric case (corresponding to $q^2=r^2$, black symbols) have been included for comparison.}
\label{fig:gamma1qp}
\end{figure}

The resulting form factor for the tree-level tensor, $\overline{\Gamma}_1(q^2,q^2,p^2)$, has been represented in Fig.~\ref{fig:gamma1qp} as a function of the momenta $q$ and $p$. The symmetric ($p^2=q^2$) and soft-gluon ($p^2=0$) cases have been added for comparison
. This plot shows a smooth dependence of the form factor along the bisectoral  kinematic configurations. Indeed, this dependence can be cast in terms of a single momentum scale, $s^2=(q^2+r^2+p^2)/2$, which is the symmetric combination of the three momenta that defines the plane in Fig.~\ref{fig:triangle}(b). This fact has been denominated \emph{planar degeneracy}~\cite{Pinto-Gomez:2022brg,Pinto-Gomez:2022qjv}. Continuum methods based in Schwinger-Dyson equations also predict the leading role played by this symmetric invariant \cite{Huber:2016tvc,Eichmann:2014xya}.

\begin{figure}[t!]
\begin{center}
    \includegraphics[width=0.8\columnwidth]{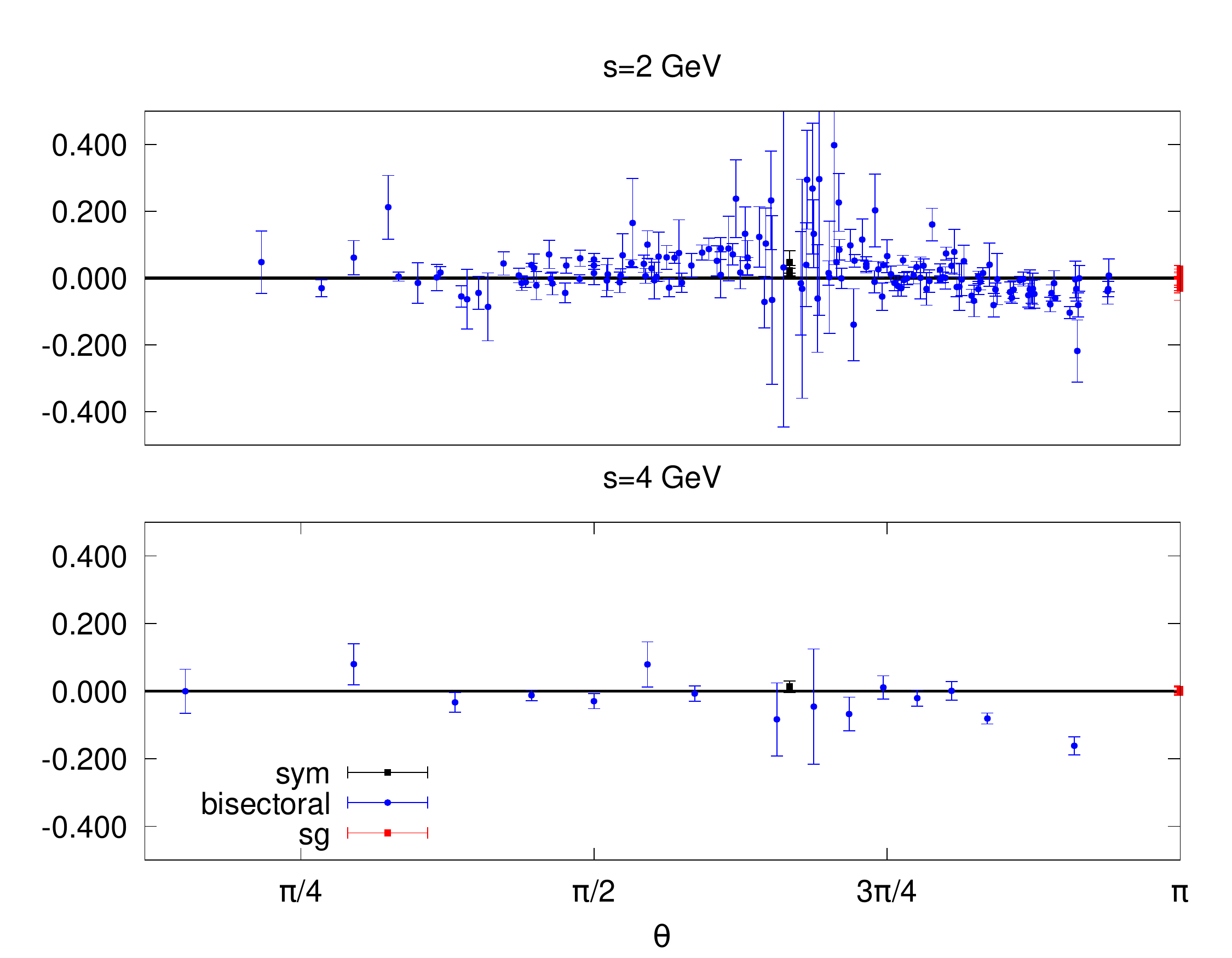}
\end{center}
\vspace{-0.75cm}
\caption{The function $d(s,\theta)$, defined in Eq.~(\ref{eq:d}), for $s=2$ GeV (top) and $s=4$ GeV (bottom) as a function of the angle between momenta $q$ and $r$ for bisectoral kinematics. 
}
\label{fig:planar}
\end{figure}

In order to deliver a deeper scrutiny of the the form factor kinematics, we have calculated its deviation with respect to the corresponding soft-gluon case,
\begin{equation}
d(s,\theta_{qr}) = \frac{\overline{\Gamma}_1(s, \theta_{qr}) - \overline{\Gamma}_{\rm sg}(s) }{\overline{\Gamma}_{\textrm{sg}}(s)} \,, \label{eq:d}
\end{equation}
along a constant $s$-curve. Considered two representative case of $s$, the outcome for $d(s,\theta_{qr})$ evaluated for all the bisectoral kinematical configurations (gray line in Fig.~\ref{fig:triangle}) is drawn in Fig.~\ref{fig:planar}. In each case, momenta have been selected in the range $(s-\delta s,s+\delta s)$, with $\delta s=0.05 {\rm GeV}$. The bisectoral configurations span over all possible angles, while the specific symmetric and soft-gluon cases ($\theta_{qr}=2\pi/3$ and $\pi$, respectively) have been highlighted for comparison. One can then conclude that deviations from the planar degeneracy appear to be small for all values of the angle $\theta_{qr}$. Namely, in very good approximation, the form factor only depends on $s$, and any dependence on other Bose-symmetric kinematic combinations can be neglected. 
In addition, mostly in the upper panel of Fig.\ref{fig:planar}, the statistical noise is seen to increase when the angle is close to the symmetric case ($\theta=2\pi/3$, black dots). As discused above, this is due to the fact that the three tensors are not independent in this limit, and the matrix inversion required for obtaining the projector $\mathcal{P}$ in Eq.(\ref{eq:projectors}) is thereby poorly conditioned. A small systematic deviation towards negative values from the planar degeneracy appears when approaching the soft-gluon limit ($\theta=\pi$, red symbols), and can be understood by relying on a simple 1-loop perturbative calculation within a massive gluon model~\cite{Pinto-Gomez:2022brg}.

\begin{figure}[t]
\begin{tabular}{cc}
\includegraphics[width=0.5\columnwidth]{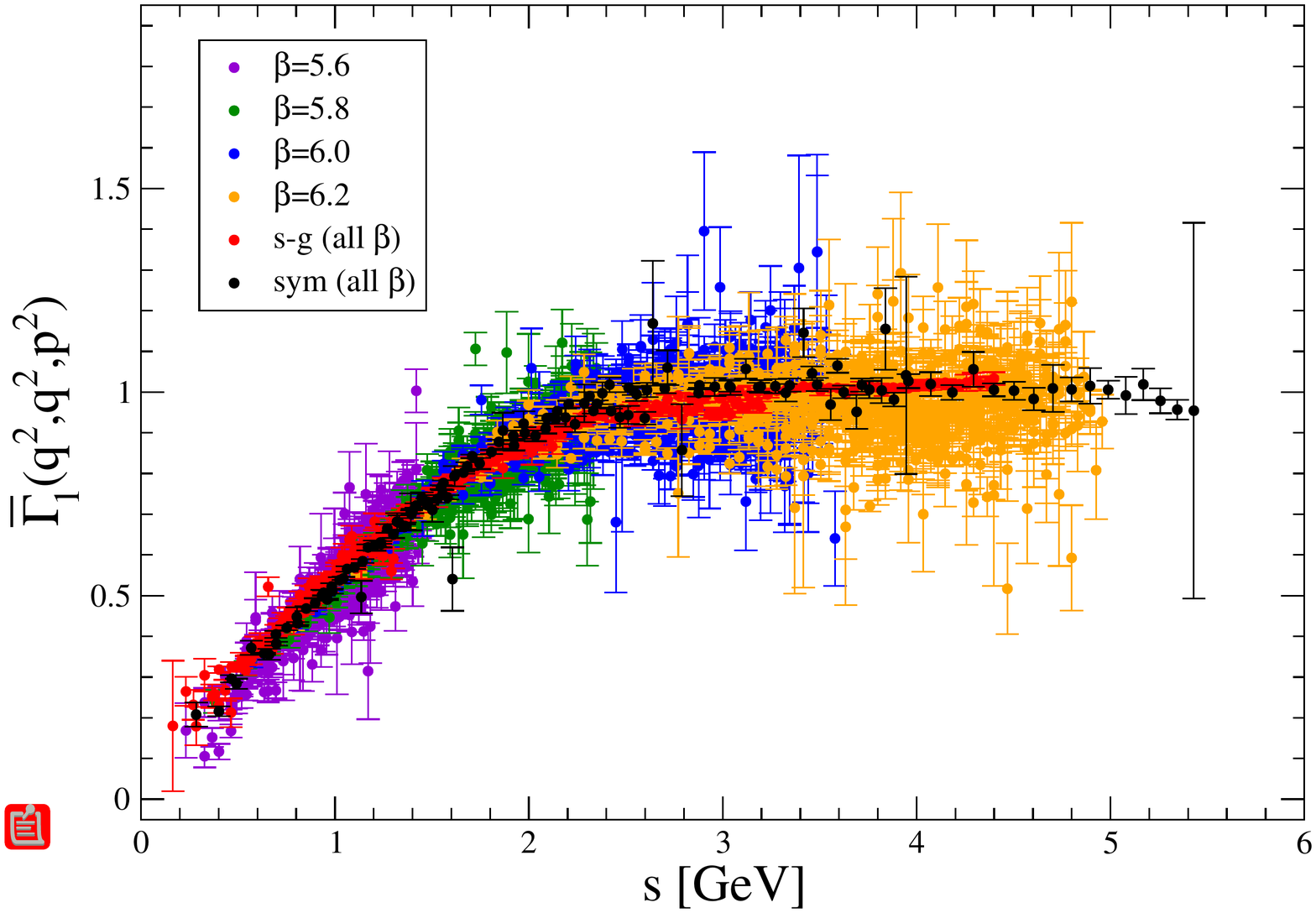}      &  \includegraphics[width=0.5\columnwidth]{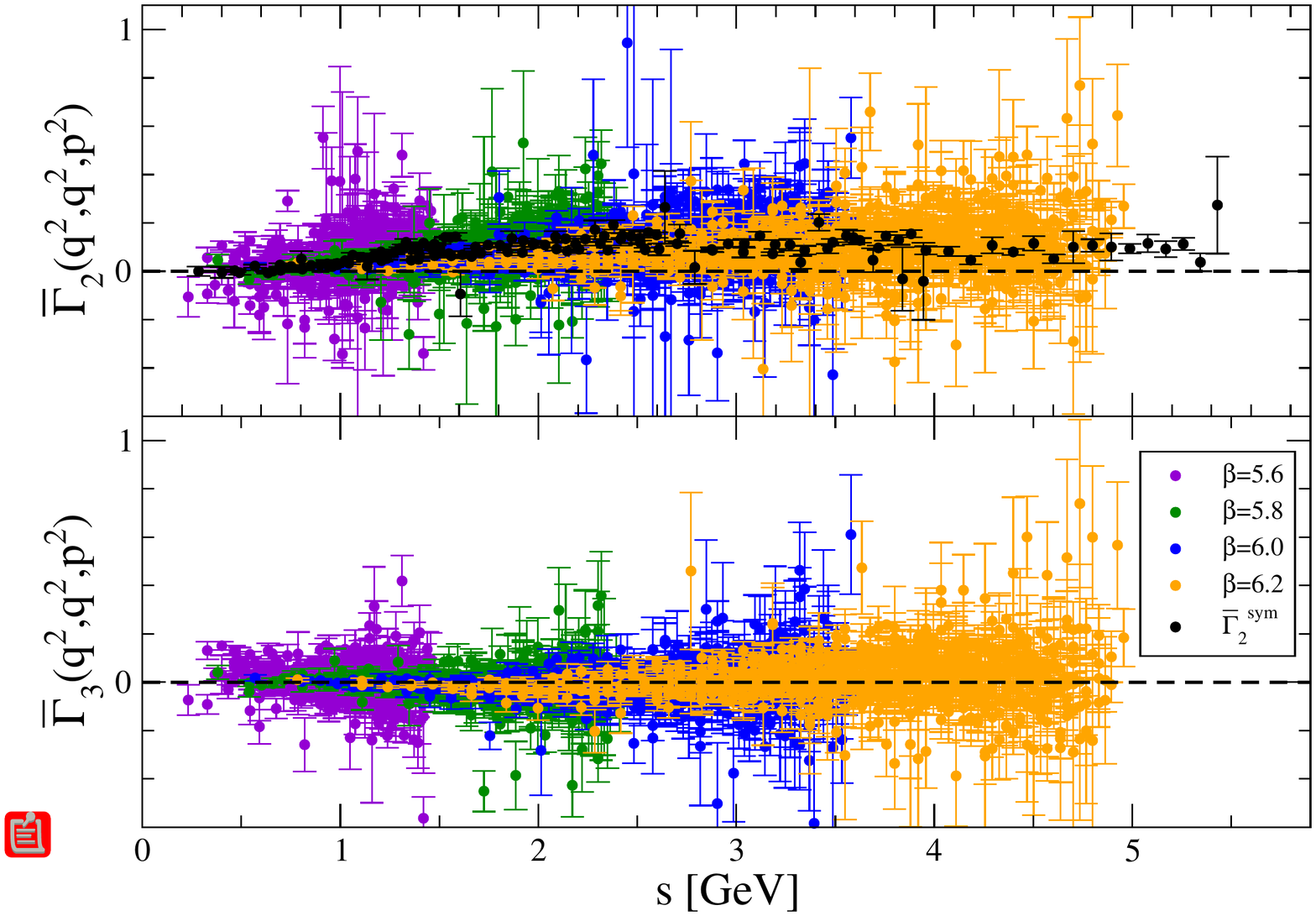} 
\end{tabular}
\vspace{-0.75cm}
\caption{Three-gluon form factors for the tree-level tensor $\lambda_1(q,r,p)$ (left), $\lambda_2(q,r,p)$ (top right), $\lambda_3(q,r,p)$ (bottom right), for bisectoral kinematic configurations $q^2=r^2$ as a function of the symmetric combination of momenta $s=\sqrt{2q^2+p^2}$. Where possible, data for the soft-gluon and symmetric cases are included for comparison.}
\label{fig:gamma123}
\end{figure}

Assuming that planar degeneracy works as a reasonable approximation, we can now plot all the three form-factors in terms of the single scale $s$ (see Fig.~\ref{fig:gamma123}). Note that these plots contain all the data points in Fig.~\ref{fig:gamma1qp} which, owing to the planar degeneracy, collapse into a single curve. This is a striking confirmation of the planar degeneracy. Furthermore, Fig.~\ref{fig:gamma123} makes also very apparent that the tree-level form factor is clearly different from zero, while the others have a subdominant role or vanish (right panel). 

A final comment is in order regarding lattice errors present in Figs.~\ref{fig:planar}-\ref{fig:gamma123}. These plots may still contain some systematic errors coming from lattice spacing or hypercubic artifacts, from which the lattice three-point Green's functions might not be plainly cured. Therefore, an application of H4-extrapolation methods~\cite{Becirevic:1999uc,Becirevic:1999hj,deSoto:2007ht,deSoto:2022scb} could have an impact on a further reduction of the statistical noise.

\section{Conclusions}
\label{sec:conclusions}

The present paper presents a non-perturbative determination of the three-gluon vertex from quenched lattice QCD 
with a significant advance with respect to previous lattice studies, that focused on very particular kinematics. Here the calculation of the vertex form factors has been extended to the case where there are two independent momentum scales. The non-perturbative determination of this vertex is of great phenomenological relevance, as it is a key ingredient for continuum methods such as Dyson-Schwinger equations for gluon or quark propagators or the quark-gluon vertex, or for the Bethe-Salpeter equations for mesons or glueballs~\cite{Souza:2019ylx,Huber:2021yfy}.

The lattice data suggest a clear dominance of the tree-level form factor that, moreover, can be written as a function of the Bose-symmetric combination of momenta $s^2=(q^2+r^2+p^2)/2$. This finding has furthermore been confirmed for general kinematics (out of the bisectoral line in Fig.~\ref{fig:triangle}) in \cite{Pinto-Gomez:2022qjv}, pointing towards a suitable description of the transversely three-gluon vertex:
\begin{equation}
    \overline\Gamma^{\alpha\mu\nu}(q,r,p) = \overline{\Gamma}_{sg}\left(\frac{q^2+r^2+p^2}{2}\right) \lambda_1^{\alpha\mu\nu}(q,r,p)\ .
\end{equation}
The structure of the transversely projected three-gluon vertex inferred from this study has been employed in \cite{Aguilar:2022thg} as lattice input for calculating the displacement of the Ward identity satisfied by the pole-free part of the three-gluon vertex $\Gamma_{\alpha\mu\nu}(q,r,p)$. This displacement shows that gluon three-point function possesses a longitudinally simple pole contribution $V_{\alpha\mu\nu}(q,r,p)$ which can be associated to a coloured massless scalar gluon correlation triggering the Schwinger mechanism for gluon mass generation in a four-dimensional Yang-Mills theory.

\smallskip
\noindent\textbf{Acknowledgments}. 
The authors are indebted to A. C. Aguilar, G. Eichmann and C. D. Roberts for fruitful discussions. 
The authors acknowledge financial support by Spanish Ministry of Science and Innovation (MICINN) (grant nos. PID2019-107844-GB-C22, PID2020-113334-GB-I00); Generalitat Valenciana (grant no. Prometeo/2019/087); and Junta de Andalucía (grant nos. P18-FR-5057, UHU-1264517).
All calculations have been performed at the UPO computing center, C3UPO.

\vspace*{-0.25cm}

\bibliographystyle{unsrt}
\bibliography{refs}

\begin{thebibliography}{10}

\bibitem{Roberts:2020hiw}
Craig~D Roberts.
\newblock {Empirical Consequences of Emergent Mass}.
\newblock {\em Symmetry}, 12(9):1468, 2020.

\bibitem{Cucchieri:2006tf}
Attilio Cucchieri, Axel Maas, and Tereza Mendes.
\newblock {Exploratory study of three-point Green's functions in Landau-gauge
  Yang-Mills theory}.
\newblock {\em Phys. Rev.}, D74:014503, 2006.

\bibitem{Cucchieri:2008qm}
Attilio Cucchieri, Axel Maas, and Tereza Mendes.
\newblock {Three-point vertices in Landau-gauge Yang-Mills theory}.
\newblock {\em Phys. Rev.}, D77:094510, 2008.

\bibitem{Huber:2016tvc}
Markus~Q. Huber.
\newblock {Correlation functions of three-dimensional Yang-Mills theory from
  Dyson-Schwinger equations}.
\newblock {\em Phys. Rev. D}, 93(8):085033, 2016.

\bibitem{Huber:2012zj}
Markus~Q. Huber, Axel Maas, and Lorenz von Smekal.
\newblock {Two- and three-point functions in two-dimensional Landau-gauge
  Yang-Mills theory: Continuum results}.
\newblock {\em JHEP}, 11:035, 2012.

\bibitem{Pelaez:2013cpa}
Marcela Pelaez, Matthieu Tissier, and Nicolas Wschebor.
\newblock {Three-point correlation functions in Yang-Mills theory}.
\newblock {\em Phys. Rev.}, D88:125003, 2013.

\bibitem{Aguilar:2013vaa}
A.~C. Aguilar, D.~Binosi, D.~Iba{\~n}ez, and J.~Papavassiliou.
\newblock {Effects of divergent ghost loops on the Green's functions of QCD}.
\newblock {\em Phys. Rev.}, D89:085008, 2014.

\bibitem{Blum:2014gna}
Adrian Blum, Markus~Q. Huber, Mario Mitter, and Lorenz von Smekal.
\newblock {Gluonic three-point correlations in pure Landau gauge QCD}.
\newblock {\em Phys. Rev.}, D89:061703, 2014.

\bibitem{Mitter:2014wpa}
Mario Mitter, Jan~M. Pawlowski, and Nils Strodthoff.
\newblock {Chiral symmetry breaking in continuum QCD}.
\newblock {\em Phys. Rev.}, D91:054035, 2015.

\bibitem{Williams:2015cvx}
Richard Williams, Christian~S. Fischer, and Walter Heupel.
\newblock {Light mesons in QCD and unquenching effects from the 3PI effective
  action}.
\newblock {\em Phys. Rev.}, D93(3):034026, 2016.

\bibitem{Blum:2015lsa}
Adrian~L. Blum, Reinhard Alkofer, Markus~Q. Huber, and Andreas Windisch.
\newblock {Unquenching the three-gluon vertex: A status report}.
\newblock {\em Acta Phys. Polon. Supp.}, 8(2):321, 2015.

\bibitem{Cyrol:2016tym}
Anton~K. Cyrol, Leonard Fister, Mario Mitter, Jan~M. Pawlowski, and Nils
  Strodthoff.
\newblock {Landau gauge Yang-Mills correlation functions}.
\newblock {\em Phys. Rev.}, D94(5):054005, 2016.

\bibitem{Duarte:2016ieu}
Anthony~G. Duarte, Orlando Oliveira, and Paulo~J. Silva.
\newblock {Further Evidence For Zero Crossing On The Three Gluon Vertex}.
\newblock {\em Phys. Rev.}, D94(7):074502, 2016.

\bibitem{Sternbeck:2017ntv}
Andre Sternbeck, Paul-Hermann Balduf, Ayse Kizilersu, Orlando Oliveira,
  Paulo~J. Silva, Jon-Ivar Skullerud, and Anthony~G. Williams.
\newblock {Triple-gluon and quark-gluon vertex from lattice QCD in Landau
  gauge}.
\newblock {\em PoS}, LATTICE2016:349, 2017.

\bibitem{Corell:2018yil}
Lukas Corell, Anton~K. Cyrol, Mario Mitter, Jan~M. Pawlowski, and Nils
  Strodthoff.
\newblock {Correlation functions of three-dimensional Yang-Mills theory from
  the FRG}.
\newblock {\em SciPost Phys.}, 5(6):066, 2018.

\bibitem{Vujinovic:2018nqc}
Milan Vujinovic and Tereza Mendes.
\newblock {Probing the tensor structure of lattice three-gluon vertex in Landau
  gauge}.
\newblock {\em Phys. Rev.}, D99(3):034501, 2019.

\bibitem{Barrios:2022hzr}
Nahuel Barrios, Marcela Pel\'aez, and Urko Reinosa.
\newblock {Two-loop three-gluon vertex from the Curci-Ferrari model and its
  leading infrared behavior to all loop orders}.
\newblock {\em arXiv:2207.10704 [hep-ph]}, 7 2022.

\bibitem{Aguilar:2019kxz}
A.C. Aguilar, M.N. Ferreira, C.T. Figueiredo, and J.~Papavassiliou.
\newblock {Gluon mass scale through nonlinearities and vertex interplay}.
\newblock {\em Phys. Rev. D}, 100(9):094039, 2019.

\bibitem{Aguilar:2021uwa}
A.~C. Aguilar, M.~N. Ferreira, and J.~Papavassiliou.
\newblock {Exploring smoking-gun signals of the Schwinger mechanism in QCD}.
\newblock {\em Phys. Rev. D}, 105(1):014030, 2022.

\bibitem{Pinto-Gomez:2022brg}
F.~Pinto-G\'omez, F.~De~Soto, M.~N. Ferreira, J.~Papavassiliou, and
  J.~Rodr\'\i{}guez-Quintero.
\newblock {Lattice three-gluon vertex in extended kinematics: planar
  degeneracy}.
\newblock 8 2022.

\bibitem{Athenodorou:2016oyh}
A.~Athenodorou, D.~Binosi, Ph. Boucaud, F.~De~Soto, J.~Papavassiliou,
  J.~Rodriguez-Quintero, and S.~Zafeiropoulos.
\newblock {On the zero crossing of the three-gluon vertex}.
\newblock {\em Phys. Lett.}, B761:444--449, 2016.

\bibitem{Boucaud:2017obn}
Ph. Boucaud, F.~De~Soto, J.~Rodríguez-Quintero, and S.~Zafeiropoulos.
\newblock {Refining the detection of the zero crossing for the three-gluon
  vertex in symmetric and asymmetric momentum subtraction schemes}.
\newblock {\em Phys. Rev.}, D95(11):114503, 2017.

\bibitem{Aguilar:2021lke}
A.~C. Aguilar, F.~De~Soto, M.~N. Ferreira, J.~Papavassiliou, and
  J.~Rodr\'\i{}guez-Quintero.
\newblock {Infrared facets of the three-gluon vertex}.
\newblock {\em Phys. Lett. B}, 818:136352, 2021.

\bibitem{Aguilar:2021okw}
A.~C. Aguilar, C.~O. Ambr\'osio, F.~De~Soto, M.~N. Ferreira, B.~M. Oliveira,
  J.~Papavassiliou, and J.~Rodr\'\i{}guez-Quintero.
\newblock {Ghost dynamics in the soft gluon limit}.
\newblock {\em Phys. Rev. D}, 104(5):054028, 2021.

\bibitem{Ibanez:2012zk}
D.~Ib\'a{\~n}ez and J.~Papavassiliou.
\newblock {Gluon mass generation in the massless bound-state formalism}.
\newblock {\em Phys. Rev.}, D87(3):034008, 2013.

\bibitem{Binosi:2017rwj}
Daniele Binosi and Joannis Papavassiliou.
\newblock {Coupled dynamics in gluon mass generation and the impact of the
  three-gluon vertex}.
\newblock {\em Phys. Rev.}, D97(5):054029, 2018.

\bibitem{Ball:1980ax}
James~S. Ball and Ting-Wai Chiu.
\newblock {Analytic properties of the vertex function in gauge theories. 2}.
\newblock {\em Phys. Rev.}, D22:2550, 1980.

\bibitem{Ball:1980ay}
James~S. Ball and Ting-Wai Chiu.
\newblock {Analytic Properties of the Vertex Function in Gauge Theories. 1.}
\newblock {\em Phys. Rev.}, D22:2542, 1980.

\bibitem{Boucaud:2018xup}
Ph. Boucaud, F.~De~Soto, K.~Raya, J.~Rodríguez-Quintero, and S.~Zafeiropoulos.
\newblock {Discretization effects on renormalized gauge-field Green’s
  functions, scale setting, and the gluon mass}.
\newblock {\em Phys. Rev.}, D98(11):114515, 2018.

\bibitem{Pinto-Gomez:2022qjv}
F.~Pinto-Gomez and F.~de~Soto.
\newblock {Three-gluon vertex in Landau-gauge from quenched-lattice QCD in
  general kinematics}.
\newblock 11 2022.

\bibitem{Eichmann:2014xya}
Gernot Eichmann, Richard Williams, Reinhard Alkofer, and Milan Vujinovic.
\newblock {The three-gluon vertex in Landau gauge}.
\newblock {\em Phys. Rev.}, D89:105014, 2014.

\bibitem{Becirevic:1999uc}
D.~Becirevic, Ph. Boucaud, J.P. Leroy, J.~Micheli, O.~Pene,
  J.~Rodriguez-Quintero, and C.~Roiesnel.
\newblock {Asymptotic behaviour of the gluon propagator from lattice {QCD}}.
\newblock {\em Phys. Rev.}, D60:094509, 1999.

\bibitem{Becirevic:1999hj}
D.~Becirevic, Ph. Boucaud, J.P. Leroy, J.~Micheli, O.~Pene,
  J.~Rodriguez-Quintero, and C.~Roiesnel.
\newblock {Asymptotic scaling of the gluon propagator on the lattice}.
\newblock {\em Phys. Rev.}, D61:114508, 2000.

\bibitem{deSoto:2007ht}
F.~de~Soto and C.~Roiesnel.
\newblock {On the reduction of hypercubic lattice artifacts}.
\newblock {\em JHEP}, 0709:007, 2007.

\bibitem{deSoto:2022scb}
F.~de~Soto.
\newblock {Restoring rotational invariance for lattice QCD propagators}.
\newblock {\em JHEP}, 10:069, 2022.

\bibitem{Souza:2019ylx}
Emanuel~V Souza, Mauricio Narciso~Ferreira, Arlene~Cristina Aguilar, Joannis
  Papavassiliou, Craig~D Roberts, and Shu-Sheng Xu.
\newblock {Pseudoscalar glueball mass: a window on three-gluon interactions}.
\newblock {\em Eur. Phys. J. A}, 56(1):25, 2020.

\bibitem{Huber:2021yfy}
Markus~Q. Huber, Christian~S. Fischer, and Helios Sanchis-Alepuz.
\newblock {Higher spin glueballs from functional methods}.
\newblock {\em Eur. Phys. J. C}, 81(12):1083, 2021.

\bibitem{Aguilar:2022thg}
A.~C. Aguilar, F.~De~Soto, M.~N. Ferreira, J.~Papavassiliou, F.~Pinto-G\'omez,
  C.~D. Roberts, and J.~Rodr\'\i{}guez-Quintero.
\newblock {Schwinger mechanism for gluons from lattice QCD}.
\newblock 11 2022.

\end{thebibliography}

\end{document}